\documentclass[]{aastex631}

\newcommand{\msfc}{NASA Marshall Space Flight Center, Huntsville, AL 35812, USA}
\newcommand{\uah}{Center for Space Plasma and Aeronomic Research, University of Alabama Huntsville, Huntsville, AL, 35812, USA}
\newcommand{\minesparis}{École des Mines de Paris, 60 Bd Saint-Michel, 75272 Paris, France}

\newcommand{\magixs}{\textit{MaGIXS}}

\newcommand{\magixsone}{MaGIXS-1}

\newcommand{\degree}{$^{\circ}$}

\begin{document}
\shorttitle{Inversion of Overlappograms}
\shortauthors{Athiray et. al}
\newcommand{\psa}[1]{{{\bf\color{blue} #1}}}
\newcommand{\amy}[1]{{{\bf\color{red} #1}}}
\newcommand{\arthur}[1]{{{\bf\color{green} #1}}}

\title{A systematic study of inverting overlappograms : MaGIXS - A case study}

 \author[0000-0002-4454-147X]{P. S. Athiray}\affiliation{\uah}\affiliation{\msfc}
\author{Arthur Hochedez}\affiliation{\uah}\affiliation{\minesparis}
\author[0000-0002-5608-531X]{Amy R. Winebarger}\affiliation{\msfc}
\author{Dyana Beabout}\affiliation{\msfc}
\begin{abstract}

Slitless (or wide field) imaging spectroscopy provides simultaneous imaging and spectral information from a wide field of view, which allows for rapid spectroscopic data collection of extended sources. Depending on the size of the extended source combined with the spatial resolution and spectral dispersion of the instrument, there may be locations in the focal plane where spectral lines from different spatial locations overlap  on the detector. An unfolding method has been successfully developed and demonstrated on the recent rocket flight of the Marshall Grazing Incidence X-ray Spectrometer (MaGIXS), which observed several strong  emission lines in the 8 to 30 \AA\ wavelength range from two X-ray bright points and a portion of an active region. In this paper, we present a systematic investigation of the parameters that control and optimize the inversion method to unfold slitless spectrograph data.

\end{abstract}

\section{Introduction} \label{sec:intro}

For the past four decades, the majority of solar imaging spectrographs have obtained, at any given time, spectral information from a narrow slit and spatial information only along the slit. The position of the slit on the Sun is often moved, called `rastering', to generate spectral information over  a wide field of view. However, rastering can be a  slow process and the resulting 2-D information is a convolution of both spatial morphological changes and temporal evolution.  Slitless or wide field imaging spectrographs can simultaneously obtain detailed spatial and spectral information of an extended object. (Note we use the term slitless and wide field interchangeably and that some wide field imaging spectrographs may use a wide slot in place of the slit.)  The data from such instruments, called `spectroheliograms' or `overlappograms', contain spectral information from different spatial locations on the Sun that can overlap in the dispersive direction. Because of these overlaps, an unfolding routine may be required to appropriately analyze the data.

The study of the solar corona via slitless imaging spectroscopy has a long history. The first overlappograms routinely obtained were from the Naval Research Laboratory (NRL) experiment, S-082A spectroheliograph, onboard Skylab \citep{tousey1973}.  This instrument observed dispersed spectral images of the Sun in the 150 - 625\,\AA\ EUV wavelength range, and demonstrated the richness of the spectral signatures for different solar features. The Res-C instrument onboard the Russian CORONAS-I mission \citep{kuzin1997,zhitnik1998} included 51 lines in the spectral region of 18.0 - 21.0 nm. This instrument was designed to achieve high spectral resolution at the expense of spatial resolution in the dispersive direction.  The Extreme-ultraviolet (EUV) Imaging Spectrometer (EIS; \citealt{culhane2007}) onboard the Hinode satellite has two wide slots (40\arcsec\ and 266\arcsec\ wide) that can also yield overlappogram data. Most recently, the Marshall Grazing Incidence X-ray Spectrometer (MaGIXS) sounding rocket experiment demonstrated successful observation of several solar coronal structures using high temperature diagnostic emission lines in the soft-X-ray wavelength range via simultaneous wide field imaging and spectroscopy \citep{champey2022,savage2022}.

Additionally, two computed tomography imaging spectrographs (CTIS) have been flown on sounding rockets.  A CTIS is an imaging spectrograph (or multiple imaging spectrographs) that captures a scene with different angles of dispersion or in different orders of light.  The Multi-Order Solar EUV Spectrograph (MOSES), a novel objective grating telescope that simultaneously captured three spectral orders of the strong He II 30.4 nm line over a field-of-view of 10\arcmin\ ${\times}20$\arcmin, was designed, built and successfully flown on a sounding rocket platform \citep{fox2010}. This instrument captured the spatial information in the zeroth order, and the spectral and spatial information in the plus and minus one orders. This instrument design also allows detection of Doppler shifts, as the spectral displacements for the plus and minus one orders are in opposite directions. The EUV Snapshot Imaging Spectrometer (ESIS) instrument is a CTIS that captured a quiet Sun scene in the He I 584\,\AA\ and O V 630\,\AA\ lines with different angles of dispersion and an 11\arcmin\ wide octagonal field of view.  ESIS was built and launched on a sounding rocket on September 30th, 2019 \citep{parker2022}.

Significant efforts have been made to develop robust unfolding methods to analyze and better interpret these data (see examples in \citealt{fox2010,harra2017,davila2019,parker2022,cheung2019,winebarger2019}).

The \magixs{} flight data has been inverted with the ElasticNet inversion routine, a standard Python inversion routine that minimizes the least squares, as well as the L1 and L2 norms of the solution \citep{savage2022,athiray2024}.   The method relies on several parameters that control how much penalty is added to the solution and the relative importance of the L1 and L2 norms.

In this paper, we present a systematic investigation of the ElasticNet inversion method, with a primary goal to assess the optimization of inversion parameters and validate the robustness of the inversion. In addition, we also demonstrate the inclusion of weights in the inversion, which helps to improve the inversion and overcome minor artifacts arising due to spatio-spectral confusion. We use the X-ray bright points observation from the first \magixs\ flight data  (here after \magixsone) as our test case for this evaluation exercise.  This paper is organized as follows.  An overview of \magixsone\ flight and data used for this work are presented in Section 2. In Section 3, we introduce the inversion method and describe the implementation of adding weights to the inversion. Optimization of parameters for the inversion using statistical metrics are described in Section 4.  Results of the inversion and summary are presented in Section 5.

\section{\magixsone\ flight overview} \label{sec:magixsoverview}

Here we present a summary of \magixsone{} observations and describe the flight data, which is used for studying the parameters of the inversion. Additional details of \magixsone{} instrument can be found in \cite{athiray2021} and \cite{champey2022} and the \magixsone{} mission, including the standard data processing techniques, in \cite{savage2022}.

\begin{figure}[h!]
 \centering
    \includegraphics[width=0.95\textwidth]{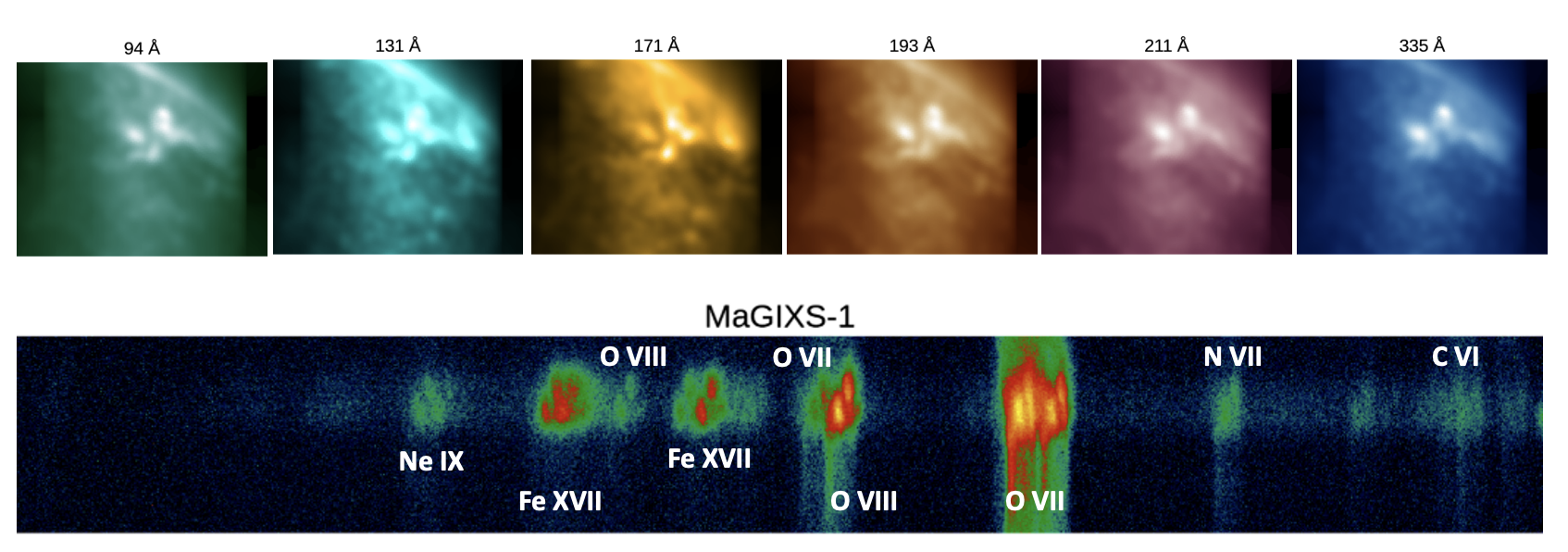}
    \caption{(Top) Time averaged AIA cutout images of the X-ray bright points corresponding to \magixsone\ observations, degraded to match \magixsone\ resolution including the vignetting function. (Bottom) \magixsone{} overlappogram data that shows spectrally dispersed  emission lines from the X-ray bright points, with strong lines identified.  }
    \label{fig:AIA_magixsfov}
\end{figure}

\magixsone{} is an X-ray imaging spectrograph designed to fly on a NASA sounding rocket to observe the Sun in the 8 to 30 \AA\ wavelength range. The primary science objective of \magixsone{} was to observationally determine the frequency of heating events in solar active region structures using several high temperature diagnostic emission lines in soft X-rays \citep{Athiray2019}. The effective field of view observed in \magixsone{} is roughly ${\sim}$ 9.25 \arcmin\ ${\times}$ 25 \arcmin, defined by both a slot at an intermediate focus position and the instrument vignetting function \citep{champey2022}.   The successful launch of \magixsone{} occurred on 30 July 2021 with data collection occurring from 18:21 to 18:26 UT.  The observed targets included two X-ray bright points on the northern hemisphere of the solar disk and a portion of an active region on the southern hemisphere of the disk. After initial pointing maneuvers, \magixsone{} observed the targets for about 296 seconds with 2 second cadence during stable pointing. In this study, we use the time averaged data collected over stable pointing. We also note that \magixsone{} flight data are modulated by the instrument vignetting function, which peaks at the center of the field of view (FOV) and decreases on either side of the field.

Figure \ref{fig:AIA_magixsfov} (top panel) shows the time averaged cutout images of the  Atmospheric Imaging Assembly (AIA; \citealt{lemen2012}) that have been degraded to \magixsone{} resolution, including applying the appropriate vignetting function. Spectrally dispersed X-ray emission from these two coronal spatial structures are observed in the \magixsone{} flight data, shown in Figure~\ref{fig:AIA_magixsfov} (bottom panel) with key spectral lines identified. The \magixsone{} Level 1.5 data of these two bright points will be utilized for this study.

\section{{\magixsone} Inversion} \label{sec:magixsinversion}

\begin{figure}[h!]
    \centering
    \includegraphics[width=0.5\textwidth]{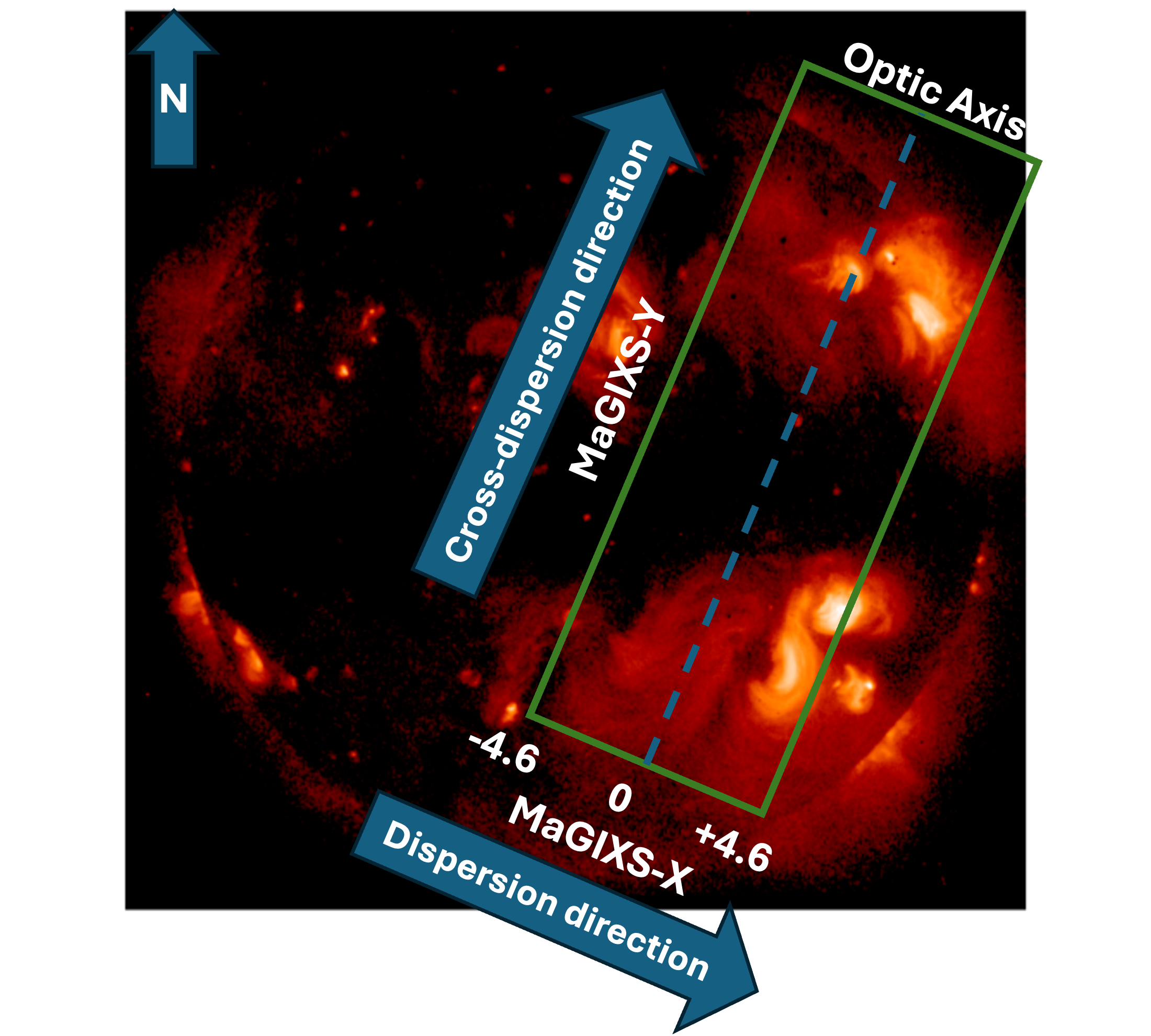}
    \caption{Pictorial representation (not drawn to scale) to define field angles for \magixsone{}. Green rectangle box defines \magixsone{} field of view (9.25 \arcmin in the dispersion direction and  25 \arcmin in the cross-dispersion directions, marked with arrows. Field angles are measured in the dispersion direction relative to the optical axis. The field angles are analogous to helio-centric coordinates but measured from the instrument optic axis.}
    \label{fig:Fieldangle}
\end{figure}

 The inversion method adopts the general framework of spectral decomposition described in \cite{cheung2019} and \cite{winebarger2019}. The problem is first written as a set of linear equations, namely
\begin{equation}
    y = Mx
    \label{eqn:lin}
\end{equation}
where $y$ is a one-dimensional array that contains a single row of the \magixsone{} flight data (2048 elements), $x$ is a one-dimensional array of emission measures at different solar locations (or field angles) and temperatures that can contribute to the flight data ($n_{free}$), and $M$ is the response matrix that contains how emission measure from each solar location and temperature map into electrons at specific pixels on the detector ($n_{free} {\times}2048)$. Here we define the term `field angle' to represent distance in the dispersion direction measured from the optical axis. Figure \ref{fig:Fieldangle} shows a graphical representation of the \magixsone{} instrument axis. \magixsone{}, was rolled relative to solar north (23 \degree counter clockwise) and pointed at a location near the limb.  The field angles  are analogous to solar-X, but measured from the optical axis of the instrument.  We refer to this direction as MaGIXS-X.  Further, we refer to distances in the cross-dispersion as MaGIXS-Y.

The response matrix is generated by first calculating a set of isothermal spectra at different temperatures at a unit emission measure ($1 {\times}10^{26}$ cm$^{-5}$) using CHIANTI atomic database \citep{Dere2023}. We included the newly added atomic calculations for the satellite lines of  Fe {\sc xvi} \citep{delzanna2024_fe16}.  These isothermal spectra are then convolved with the effective area measured from \magixsone{} ground calibration \citep{athiray2020, athiray2021}.  Finally, the response from each unique field angle is mapped into the detector using the wavelength calibration determined from flight data \citep{savage2022}. Each column of the response matrix, then, corresponds to the unique signal expected from a unit emission measure at a single temperature and field angle.

When generating the isothermal spectra, several assumptions must be made concerning the atomic parameters, including the elemental abundances, ionization state, plasma density, electron speed distribution, and ionization balance.  The inversion can be performed with different response functions that vary these assumptions to see which best match the observations.
For all the analysis presented in this paper, we used the response function generated with coronal abundances, Maxwellian electron distribution, and an assumption of ionization equilibrium.  These assumptions were found to be the best match of the flight data in \cite{savage2022}.

\begin{figure}
    \centering
     \includegraphics[width=0.9\textwidth]{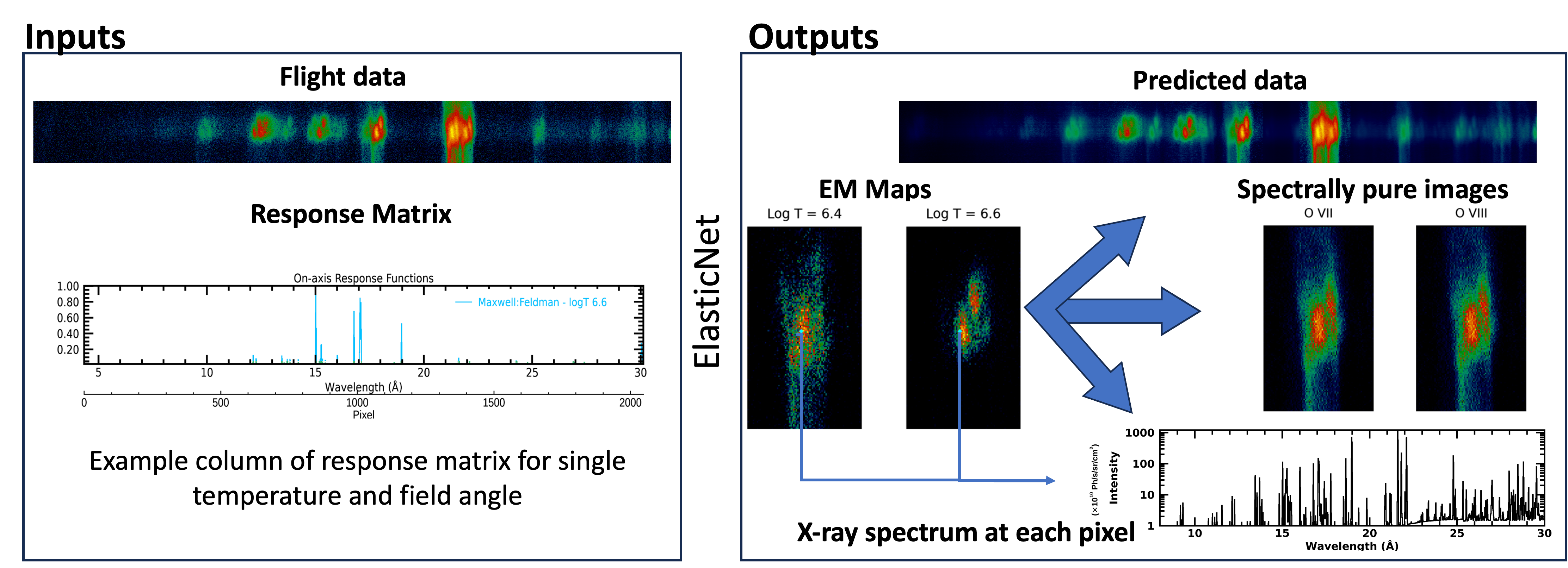}
    \caption{A graphical representation of the inputs and outputs to the inversion.  Inputs are the flight data and response matrix.  The direct output is the emission measure as a function of temperature and field angle.  Each row of the data is inverted separately, then the emission measures are combined to make emission measure maps. From the emission measure maps, we calculate the predicted data, spectrally pure maps, and spectrum at each spatial location. }
    \label{fig:thematic_rep}
\end{figure}

Additionally, when generating the response function, we establish the temperature and spatial grid on which the solution is calculated.  Ostensibly, the range of temperatures (minimum, maximum, and temperature step size in logarithmic scale) and range of field angles (minimum, maximum, and field angle grid size) are all free parameters.  However, they are all informed by the available constraints in the data.  For instance, the wavelength range of \magixsone{} includes spectral lines formed at temperature above Log T = 5.9. Though it is possible to include smaller values of Log T in the response function, there are no adequate constraints for those values.  Their inclusion would imply the inversion would take longer to complete and the results in those temperature bins would not be well constrained (see for instance \citealt{athiray2024}).  Although \magixsone\ is sensitive to measure high temperature emission lines with Log T $>$ 7.0, the bright points under study are relatively cool and did not show significant Fe XVIII emission. Therefore, we considered the maximum temperature for inversion to be Log T = 7.0.

Similarly, \magixsone{}  has an effective slot size of 9.25\arcmin\ (${\pm}4.6$ \arcmin from the optical axis) in the dispersion direction.  There is no information on locations outside of this field of view, so the minimum and maximum field angles considered in this study are ${\pm}4.6$\arcmin.  The field angle grid size, ${\delta}{\theta}$, of the response function (and resulting spatial resolution of the emission measure array) is more difficult to quantize.  The spatial resolution of the inversion should not be less than the plate scale of the observations.  In the cross-dispersion direction, \magixsone{} has a plate scale of 2.8\arcsec/pixel, but in the dispersion direction, the plate scale ranges from ${\sim}$ 5.5\arcsec\ to 9\arcsec.  However, the point spread function of the instrument is 27\arcsec\ \citep{champey2022}, and there is limited information at spatial scales smaller than this value in the flight data.

For this study, we select the spatial resolution of the response matrix to be  8.4\arcsec.  We revisit this decision in the Discussion Section (see Section \ref{sec:summary}).  The ranges and resolution of the response matrix are given in Table~\ref{tab:parameter}.  The number of free parameters, then, is $n_{free}$ = 796.

\begin{table}[h!]
    \centering
    \caption{Range and Resolution of Response Function}\label{tab:parameter}
    \begin{tabular}{|c|c|}
    \hline
    Parameter & Minimum, Maximum, Resolution\\
    \hline
         Log T &  5.9, 7.0, 0.1  \\
         Field Angles$^*$ & {-4.6\arcmin, 4.6\arcmin,} 8.4\arcsec \\  \hline
    \end{tabular}\\
    \label{tab:response_params}
    $^*$ Measured from the center of the \magixsone{} optical axis.
\end{table}

 The inversion is then completed using the ElasticNet regularization technique \citep{zou2005}, a commonly used algorithm available in Scikit-learn via Python library.  It combines two types of regularization, namely the sparsity and smoothness, while finding convergence to the best solution, which is defined by the equation:
\begin{equation}
    x^{\#} = \frac{1}{2n_{data}}argmin\left[W(||(y-Mx)||^2_{2} + {\alpha}{\rho}||x||_{1} + 0.5{\alpha}(1-{\rho})||x||^2_2)\right].
    \label{eq:soln}
\end{equation}
 The first term in Equation~\ref{eq:soln} denotes the standard weighted least squares that minimizes the deviation between the observations ($y$) and the predicted values ($Mx$). The weights applied, $W$, are discussed in detail below.  The second term is the $L_1$ regularization term, minimizing this term favors a sparse solution. The third term is the $L_2$ regularization term, minimizing this term favors a smooth solution.  There are two free parameters in Equation~\ref{eq:soln}; ${\alpha}$ controls the degree of penalty to apply to the L1 and L2 terms and ${\rho}$ controls the trade-off between the L1 and L2 regularization.

Figure \ref{fig:thematic_rep} shows a graphical representation of the inversion method.  The inputs to the inversion are the \magixsone{} overlappogram data and the response matrix. The inversion is run on each row of \magixsone{} data separately, which corresponds to a single MaGIXS-Y value.
 The output of the inversion is an array of emission measures as a function of field angles and Log T.  By completing the inversion on each row, we build up the emission measure in the MaGIXS-Y direction.
After these emission measure cubes are calculated, we then generate spectrally pure maps in different spectral lines as well as the predicted data ($Mx$).  We can further calculate additional data products, such as the spectrum at each location.

We highlight that weights are now included in the inversion to incorporate realistic pixel uncertainties; weights were not included in \cite{savage2022}.  We  implement weights derived from the uncertainty in the number of electrons observed in each pixel. The number of incident solar photons, the CCD read noise, and the noise from scattered light contributes to he uncertainty.  We estimate photon uncertainty, that is, shot noise ${\sigma}^{pix}_{shot}$, in each pixel by converting electrons to photons using on-axis pixel wavelengths. The measured read noise  ${\sigma}_{read}$ of \magixsone{} CCD is 7.3 e$^-$rms/readout. We estimated scattered photon noise from flight data, ${\sigma}_{sc}$  = ${\sim}$ 5 e$^-$ rms/readout, using pixels that do not contain bright coronal structures, and that are beyond the solar limb in the \magixsone{} flight data.  In Equation~\ref{eq:soln}, $W$ is a vector, where the values are $1/{\sigma}^{pix}$. ${\sigma}^{pix}$ is the quadrature sum of read, photon, and systematic noise, calculated as :
\begin{eqnarray}
w^{pix} & = & \frac{1}{{\sigma}^{pix}} \\
{\sigma}^{pix} & = & \sqrt{({\sigma}^{pix}_{shot})^2 +{\sigma}_{read}^2+ {\sigma}_{sc}^2}
\label{eq:error}
\end{eqnarray}

\begin{figure}
\includegraphics[width=0.57\textwidth]{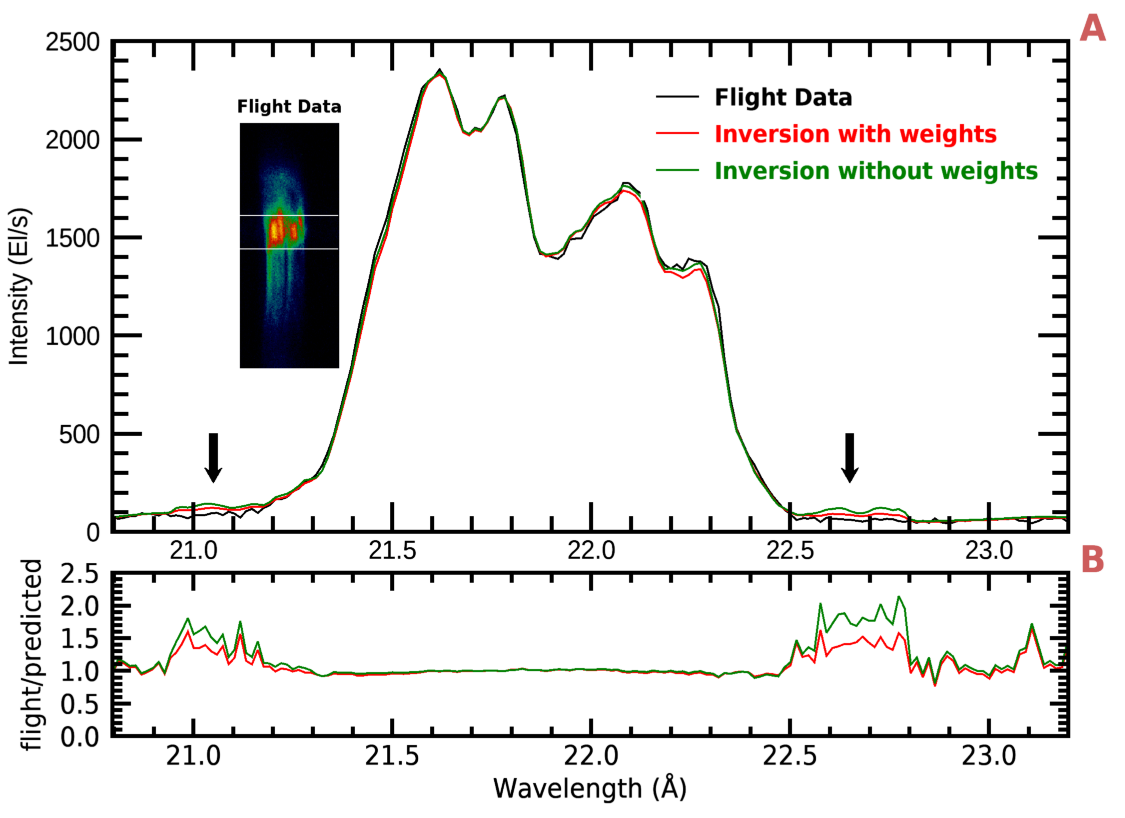}
\includegraphics[width=0.43\textwidth]{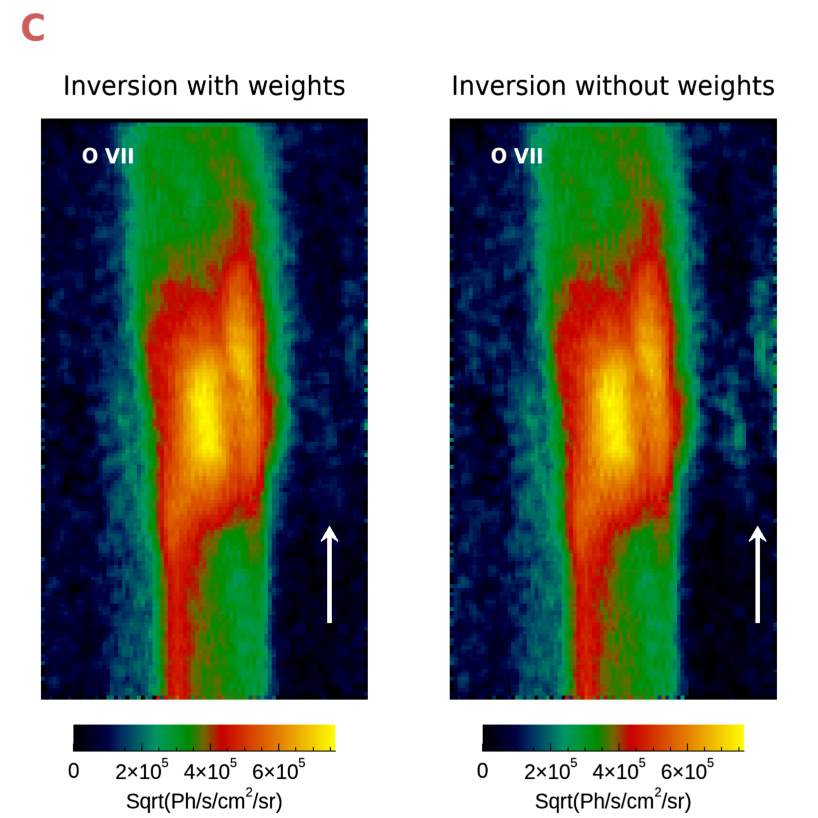}
    \caption{{Left top panel, labeled A, shows the comparison of summed \magixsone{} flight data (between two yellow horizontal lines in the inset image) with the corresponding predicted data with weights (red) and without weights (green). The bottom panel, labeled B, shows the ratio between flight data and the predicted data denoted by the respective colors. Right panel, labeled C, compare the spectrally pure images of O {\sc vii} determined from the inversion with and without weights. The white arrow indicate pixels where the artifact of ghost spatial features appear in the image when inverted without weights.}}
    \label{fig:weights}
\end{figure}

Adding weights to the inversion results in subtle but important changes in the solution.  In Figure~\ref{fig:weights} we  perform an example inversion with and without weights and compare the predicted data to the flight data around the location on the detector where the O\,{\sc vii} triplet spectral lines at 21.601, 21.802 and 22.101 \AA\ appear. The inset shows a cropped section of the flight data including the bright points in the 20.2 \AA\, to 23 \AA\, wavelength range. The flight spectra summed between the two horizontal lines in the inset image is shown in black line in Figure~\ref{fig:weights} (Left), marked as A. Overplotted are the corresponding predicted data calculated from inversions completed with weights (red) and without weights (green). The bottom panel, marked as B, shows the ratio of flight-to-predicted data for both the cases.
When weights are not included, the inversion prioritizes matching the brightest pixels on each row and can put additional emission in dimmer locations.  This can result in `ghost spatial features' highlighted with the arrows in Figure~\ref{fig:weights}; the solution with weights suppresses this excess emission.

The images in the right panel of Figure~\ref{fig:weights}, marked as C, compare the spectrally pure maps of O\,{\sc vii} determined using the inverted EM cube with and without weights.
The ghost spatial features in the inversion, denoted by white arrows,  can be seen at the edges of  the spectrally pure O\,{\sc vii} images.    From Figure \ref{fig:weights} we infer that addition of weights to the inversion better matches the observations overall so all pixels are matched within their errors, and minimizes spatial-spectral confusion artifacts.

\section{Parameter Optimization} \label{sec:parameter}

In this section, we present the methodology adopted to determine the best values for the parameters for the inversion.
 It is essential to benchmark the influence of these parameters on the inverted solution to understand the robustness of the solution. There are two parameters involved in the inversion, the L1 and L2 regularization coefficients, ${\alpha}$ and ${\rho}$, introduced in Equation~\ref{eq:soln}.

\subsection{Criteria for evaluating the free parameters}

Before we investigate the impact of the free parameters on the inversion, we first set criteria for what we anticipate would be acceptable solutions.  First, we expect every row of the inversion to converge with a reasonable number of maximum iterations.  For this study, we set the maximum number of iterations to $10^5$.  Second, we expect the predicted data ($Mx$) to well match the observed data ($y$).  As discussed in detail below, we use the centered root mean square error (RMSE)
between the observed and predicted data, Pearson’s correlation coefficient, and the standard deviation of the observed and predicted data to evaluate how well the inversion matches the observations.  Finally, we expect the spatial structures in the returned spectrally pure maps to be smooth, at least to the scale of the point spread function of the instrument.

\subsection{Establishing the Parameter Space}

The value of the penalty term, ${\alpha}$, can be any positive number.  Practically, large ${\alpha}$ does not provide a solution that matches the observations well, as the penalty terms in Equation~\ref{eq:soln} become more important than minimizing $y-Mx$.  We set the maximum ${\alpha}$ for this study to 0.05.  As ${\alpha}$ goes to 0, the inversion not only takes longer, but has rows that do not converge.  We choose the minimum value of ${\alpha}$ to be $5{\times}10^{-5}$ to avoid non-convergence.   The allowed range of ${\rho}$ is from 0 to 1, but we find ${\rho}= 0$ also has rows that do not converge.  Hence, we explore values of ${\rho}$ of 0.1 - 1.  Table~\ref{tab:free_parameters} lists the values of ${\alpha}$ and ${\rho}$ explored in this investigation.
For each combination of ${\alpha}$ and ${\rho}$, we perform an inversion and evaluate the results.
There are a total of 24 inversions considered in this study.

\begin{table}[h!]
    \centering
    \caption{List of free parameters values investigated.}
    \begin{tabular}{|c|c|}
    \hline
    Parameter & Values\\
    \hline
         ${\alpha}$ & 0.05, 5 ${\times}10^{-3}$, 5 ${\times}10^{-4}$, 5 ${\times}10^{-5}$  \\
         ${\rho}$ & 0.1, 0.3, 0.5, 0.8, 0.9, 1 \\
         \hline
         Total number of parameter combinations & 24 \\
         \hline
    \end{tabular}
    \label{tab:free_parameters}
\end{table}

\begin{figure}[h!]
    \centering
    \includegraphics[width=0.99\textwidth]{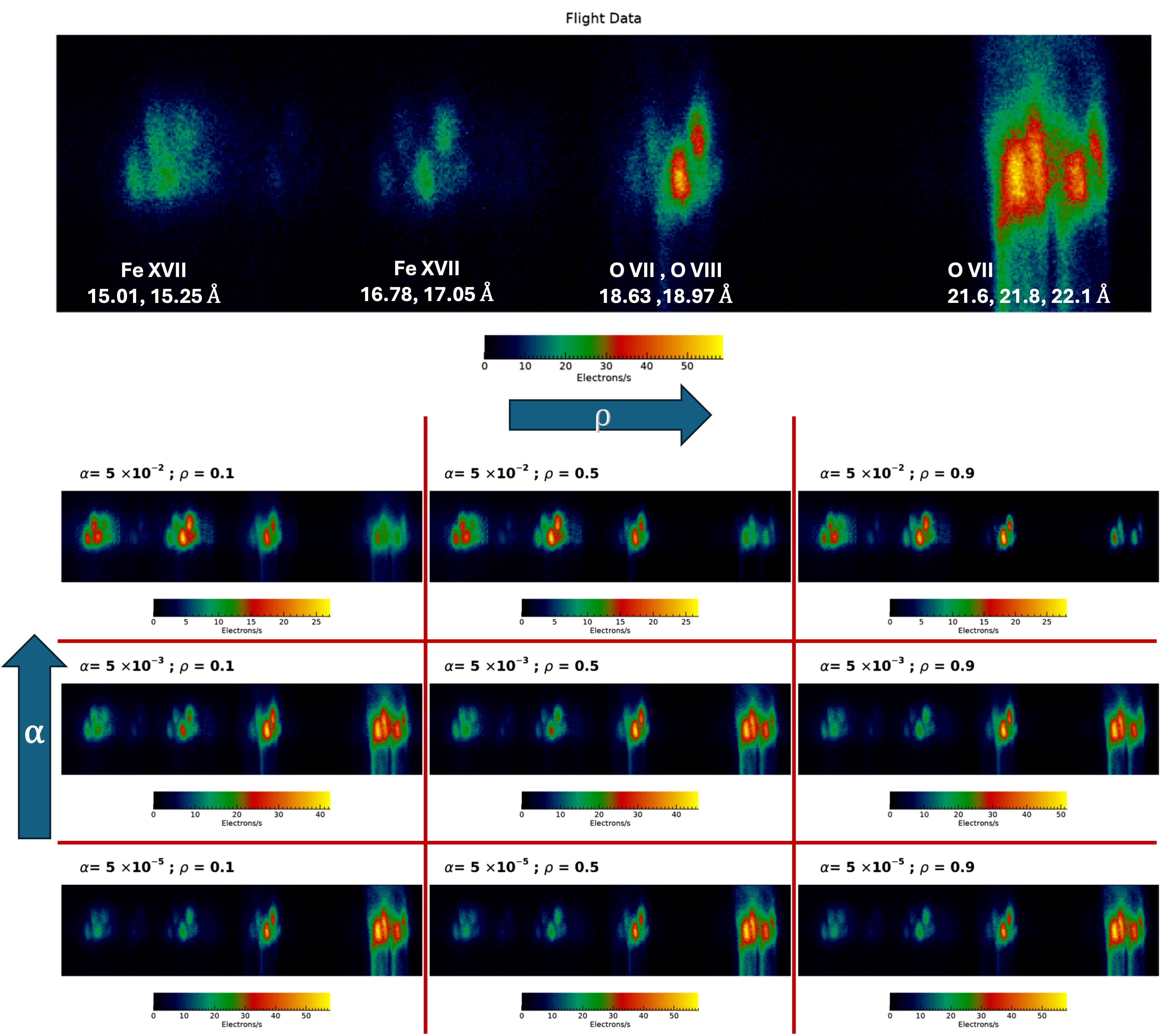}
    \caption{Comparison of flight data with predicted data for inversions with different parameter values for ${\alpha}$ and ${\rho}$. (Top) Cropped portion of \magixsone{} flight data including strong emission lines of Fe\,{\sc xvii}, O\,{\sc viii} and O\,{\sc vii} from the X-ray bright points. (Bottom) Grid of corresponding predicted data from the inversion with different combination of ${\alpha}$ and ${\rho}$ values. This plot portrays that inversion with high ${\alpha}$ (high penalty) do not match the intensity across all emission lines in the flight data. Further, high ${\alpha}$ combined with high ${\rho}$  show the effect of sparseness in the predicted data resulting in apparently a relatively compact spatial structure. We also point out the effect of the change in ${\rho}$ affecting the spatial distribution is evident at high ${\alpha}$. }
    \label{fig:spec_flight_vs_models}
\end{figure}

\subsection{Comparing Predicted and Flight Observations}

Each predicted spectroheliogram is systematically compared against the flight data to evaluate the performance of the inversion with the set of parameters. Figure \ref{fig:spec_flight_vs_models} shows a sample comparison of flight data with several inverted solutions for different ${\alpha}$ and ${\rho}$ values. The top panel in Figure  \ref{fig:spec_flight_vs_models} shows the cropped portion of the flight data that includes the strong emission lines from the X-ray bright points. The bottom panels show a sample of predicted data obtained from inversion with different ${\alpha}$ and ${\rho}$ combinations. This figure demonstrates that the inversions completed with low ${\alpha}$ values match both the spatial features at different emission lines as well as spectral intensities qualitatively.

An inversion run with an acceptable set of parameters is expected to minimize the overall deviation between the predicted and flight data. To assess the whether this is achieved, we generate a statistical summary of how well each of the predicted data match the flight data.

We employ Taylor's diagram \citep{taylor2001} to present statistical comparison of  flight data against all the predicted data. The Taylor diagram was designed as a useful graphical summary of how closely different models matches observations represented through a polar plot.  It characterizes the statistical relationship between a reference observation and test model data, and compare them in terms of their correlation, variability, and bias. These diagrams are very helpful in evaluating multiple aspects of complex models and offer relative skill scores to eliminate outlier models and identify closely matching models. Each point in a Taylor diagram represent three different statistics simultaneously, namely the centered RMSE, Pearson's correlation  coefficient, and the standard deviation (see
\citealt{taylor2001} for further details).

\begin{figure}[h!]
\includegraphics[width=0.5\textwidth]{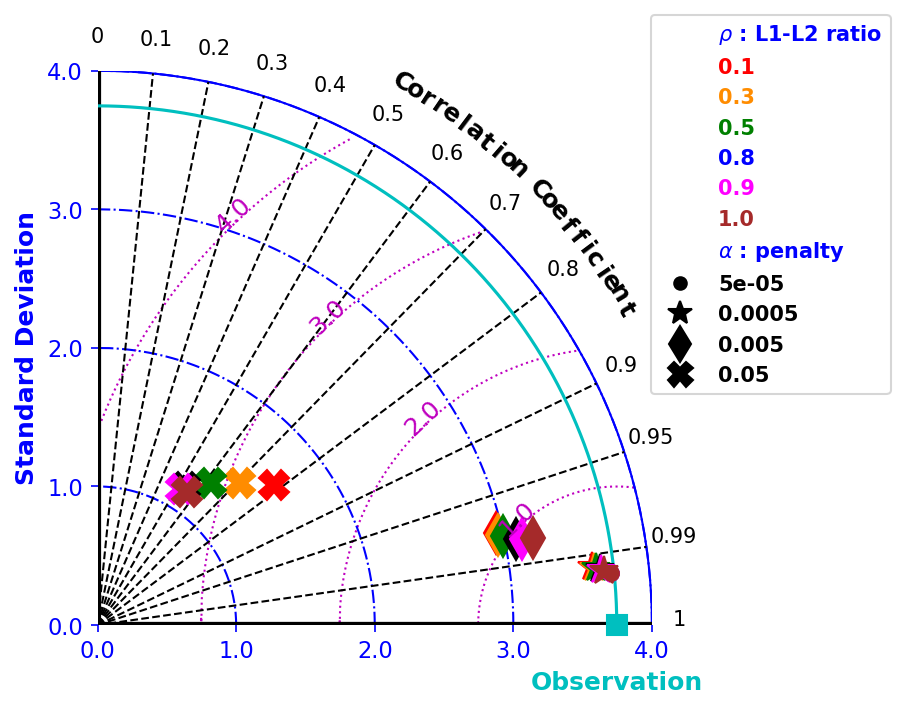}
\includegraphics[width=0.5\textwidth]{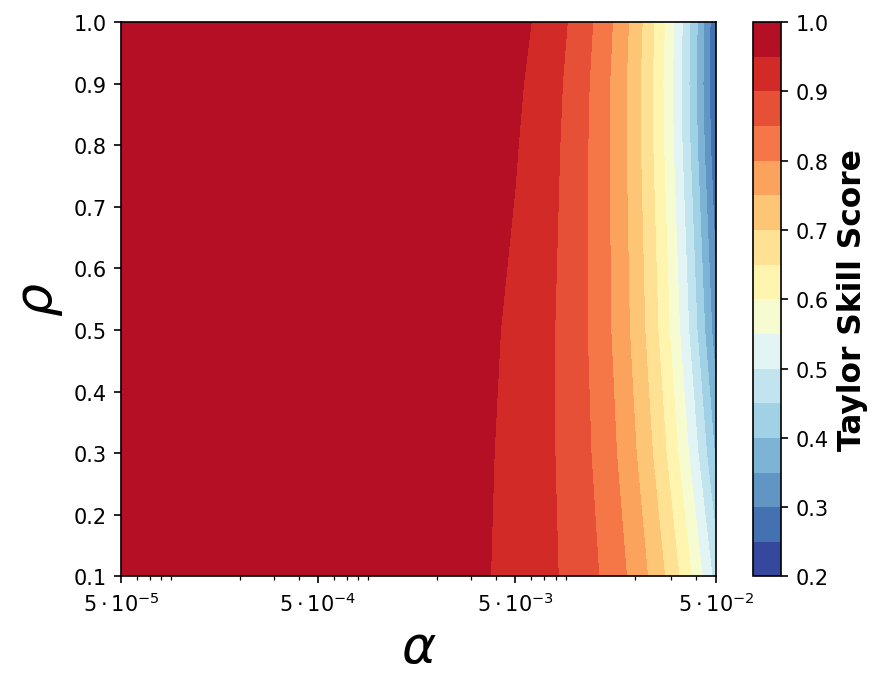}

    \caption{(Left) Taylor diagram representing a summary of different validation metrics, including the standard deviation, RMS difference, and Pearson correlation coefficient. This plot compares the predictive capabilities of models with different ${\alpha}$ (varying shapes) and ${\rho}$ (varying shape) values, with the measured \magixsone{} data.  (Right) Contour map of Taylor Skill Score varying as a function of ${\alpha}$ and ${\rho}$, with the color bar denotes the score.}
    \label{fig:taylor1}
\end{figure}

Similarly, we calculate these three values for each set of free parameters and plot them on a Taylor's diagram.  By this method, we can compare different inversions with several combinations of ${\alpha}$ and ${\rho}$ and trace the impact. The left panel in Figure~\ref{fig:taylor1} shows these metrics for different values of ${\alpha}$ and ${\rho}$.  The standard deviation of the predicted data is plotted radially, while the solid cyan line is the standard deviation of the observed data.  The Pearson correlation coefficient is plotted in the theta direction.  A value of 1 would imply perfect correlation with the data.  The cyan square on the radial axis indicates this correlation.  Finally, the concentric circles centered on this X indicates the centered RMSE.  If the predicted data point falls within the closest concentric circle, the RMS difference is $<1$.  Any set of parameters that have predicted data that fall within this circle acceptably matches the observed data by all three metrics.  This Taylor's diagram shows that predicted data with ${\alpha}= 5{\times}10^{-2}$ do not match the observations, while   $5{\times}10^{-5}$ very closely match the observation.  An ${\alpha}$ value less than $5 {\times}10^{-3}$ produces acceptable metrics for some values of ${\rho}$, while other values do not.

For each set of free parameters, we also calculate the Taylor skill score, $S$, which is defined as
\begin{equation}
    S = \frac{4(1+R)}{(\hat{{\sigma}_n}+1/\hat{{\sigma}_n})^2(1+R_0)}
    \label{taylorskillscore}
\end{equation}
where $\hat{{\sigma}_n}$ is the ratio of standard deviations of the predicted and observed data, $R$ is the correlation coefficient, and $R_0$ is the maximum correlation expected (here set to 1). It is evident that when model variance approaches observed variance, and $R$ is equal to $R_0$, the score ($S$) goes to 1.

We show the Taylor skill score as a function of ${\alpha}$ and ${\rho}$ in the right panel of Figure~\ref{fig:taylor1}. The color bar shows the skill score.  This plot demonstrates that ${\alpha}= 5 {\times}10^{-5}$ and $5{\times}10^{-4}$ for all values of ${\rho}$ have a skill score $> 0.9$.  For ${\alpha}= 5{\times}10^{-3}$, ${\rho}= 1$ has a better skill score  than smaller ${\rho}$.

\subsection{Spatial Resolution of the Inversion Results}

The final criteria to select appropriate parameters is that the the spatial structures in the returned spectrally pure maps to be smooth on the expected scales by the point spread function of the instrument.  From Figure \ref{fig:taylor1}, we observe that low ${\alpha}$ yields high Taylor score irrespective of ${\rho}$ value. However, ${\rho}$ controls the smoothness and sparseness of the inverted solution, so not all values of ${\rho}$ will necessarily produce the required smoothness spectrally pure maps.

\begin{figure}
    \centering
    \includegraphics[width=0.85\textwidth]{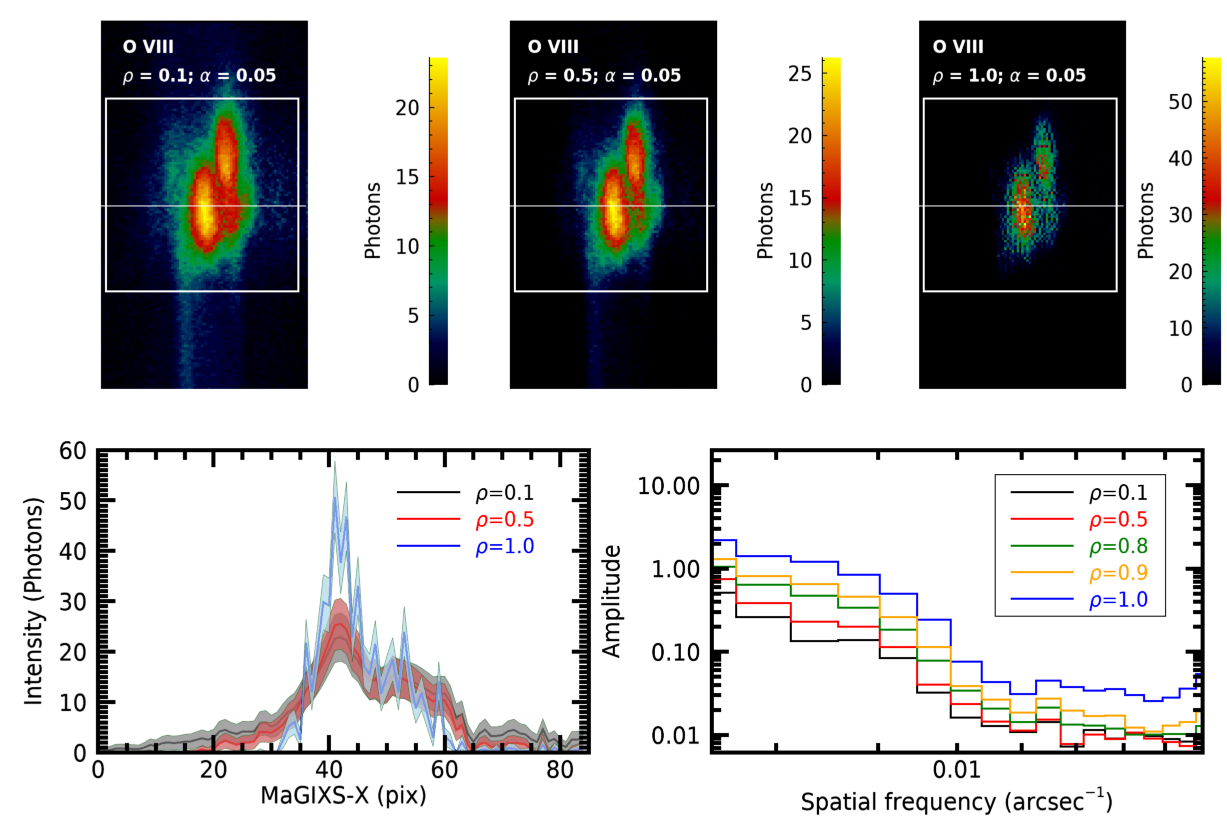}
    \caption{(Top) Comparison of {\sc O viii} maps inverted with low ${\rho}$ (smooth) and high ${\rho}$ (sparse) for a fixed ${\alpha}$ = 0.05. Inversion with high ${\rho}$ puts large amount of emission in alternate pixels and leave intermediate pixels with low emission, resulting in a ragged intensity distribution (bottom left) across the coronal structure, plotted with 1 ${\sigma}$ Poisson uncertainty shown as filled colors. The bottom right panel compares the power spectrum obtained from the selected region of interest, white square in the spectrally pure images (top panel), which shows large ${\rho}$ values result in high frequency spatial structures.}
    \label{fig:rhofinding}
\end{figure}

Figure \ref{fig:rhofinding} (Top) shows the comparison of {\sc O viii} spectrally pure
maps inverted using ${\rho}$ = 0.1, 0.5, and 1. For high ${\rho}$, the inversion favours a  sparser solution, wherein it puts large amount of emission in alternate pixels leaving the intermediate pixels with low emission. This creates intensity fluctuations in adjacent pixels, which is merely an artifact from the inversion and therefore should not be interpreted as real fluctuations on the coronal structure. Figure \ref{fig:rhofinding} (bottom left) compares the intensity distribution across the bright points (marked by white horizontal line in top panels),

Additionally, we performed Fourier analysis on the inverted spectrally pure images with different ${\rho}$ values by choosing some spatial regions encompassing the X-ray bright points, denoted by the white square in Figure \ref{fig:rhofinding} (top panels).

The bottom right  panel of Figure \ref{fig:rhofinding} is a log–log plot of such spectra for several elements of the image. The plot conveys that inversion with different ${\rho}$'s agree well in matching high frequency spatial structures greater than the  PSF of \magixsone{}. While at lower spatial frequencies higher ${\rho}$'s $>$ 0.5 introduce additional noise (artifacts) in the inverted image.  We also report that 0.1 $<$ ${\rho}$ $<$ 0.5 result in similar solutions in terms of both photon intensity and spatial distribution.

\subsection{Optimal Parameters}

Based on this study, we conclude that ${\alpha}< 5{\times}10^{-3}$ and 0.1 $<$ ${\rho}$ $<$ 0.5 produce solutions with no discernible differences.  Because smaller ${\alpha}$ require a longer run time to converge, we opt for using the largest acceptable ${\alpha}$ possible, so ${\alpha}= 5{\times}10^{-4}$.  Additionally, we choose ${\rho}= 0.5$ as the optimal ${\rho}$ value.

\section{Discussion and Conclusions} \label{sec:summary}

Slitless or wide field spectroscopy offers the exciting capability of capturing spectral information over large fields of view at a rapid cadence. However, the cost of this advancement is the need to unfold the data to remove spatial-spectral confusions.
Care must be given to select the appropriate inversion technique and parameters and to understand the impact of those choices on the resulting scientific conclusions.

In this paper, we have completed a systematic study of the weighted ElasticNet inversion method used for unfolding spatial-spectral overlaps from the spectroheliogram data observed by the \magixsone{} sounding rocket mission.  In particular, we have focused our investigation on the inversion of X-ray bright points observed in \magixsone{} data.  Unlike \cite{savage2022}, we have included weights in the inversion in this study \cite[also shown in][]{athiray2024}.  We found that the addition of uncertainties to each pixel  as weights helps the inversion to mitigate the artifact of `ghost spatial features'. This artifact arises when bright and dim spectral pixels are weighted equally in the inversion. We showed that introducing pixel weights, as determined from estimated photon noise, read noise, and scattered photon noise, significantly reduces the presence of these features and better matches the data in dim regions. Inclusion of pixel weights also enable a more powerful combined inversion of overlappogram data with imaging data, which are an order of magnitude brighter than the spectral lines in overlappogram data \citep[see for instance][]{athiray2024}.

We have carried out inversions of the \magixsone{} data with a range of parameters and statistically evaluated their performance by comparing the results to three criteria: 1) the inversion must converge with less than 10$^{5}$ iterations, 2) the predicted data must match the flight data to within the uncertainties, and 3) the smoothness of the returned spectrally pure maps must be consistent with the instrumental point spread function.

 We have carried out a detailed parameter search to determine optimum values for the inversion parameters ${\alpha}$ and ${\rho}$. We performed inversion of \magixsone{} data with a range of parameter values listed in Table~\ref{tab:free_parameters}; the ranges of these free parameters are chosen to eliminate non-convergences. We then compared the predicted data from each set of parameters with the flight data in terms of their correlation, their centered root-mean-square difference, and the ratio of their variances via Taylor diagram. We also calculated the Taylor skill score to determine optimum inversion parameters.  We found that the choice of ${\alpha}$ significantly impacts the ability for the predicted data to well match flight data, and determined ${\alpha}= 5 {\times}10^{-4}$ is the best choice to invert \magixsone{} data.  Finally, we assessed the smoothness of the spectrally pure maps resulting from each set of parameters and determined that
 0.1 ${\leq}$ ${\rho}$ ${\leq}$ 0.5, yield similar results within uncertainty limits. We choose ${\rho}= 0.5$ to be the optimum value.  Ultimately, we determine the inversion and results are insensitive to these free parameters within these specified ranges.

 When establishing the response matrix, we chose the ranges and resolutions of Log T and MaGIXS-X that would be included in the inversion.  The choices for Log T are straight forward. The coolest line \magixsone{} observed forms at Log T $>$ 5.9, which sets the lower limit of temperature for the inversion.  Non-flaring active regions are not expected to have temperatures greater than $10^7$ K, so that established the maximum temperature included in the inversion.  The shape of the emissivity functions of the spectral lines are smooth and vary on temperatures scales $>$ 0.1 in Log T, which set the resolution of temperature.  Similarly, the effective field of the view of the instrument established the minimum and maximum MaGIXS-X considered, as \magixsone{} only sees ${\pm}$ 4.6\arcmin\ from the optical axis.

\begin{figure}
    \centering
    \includegraphics[width=0.8\textwidth]{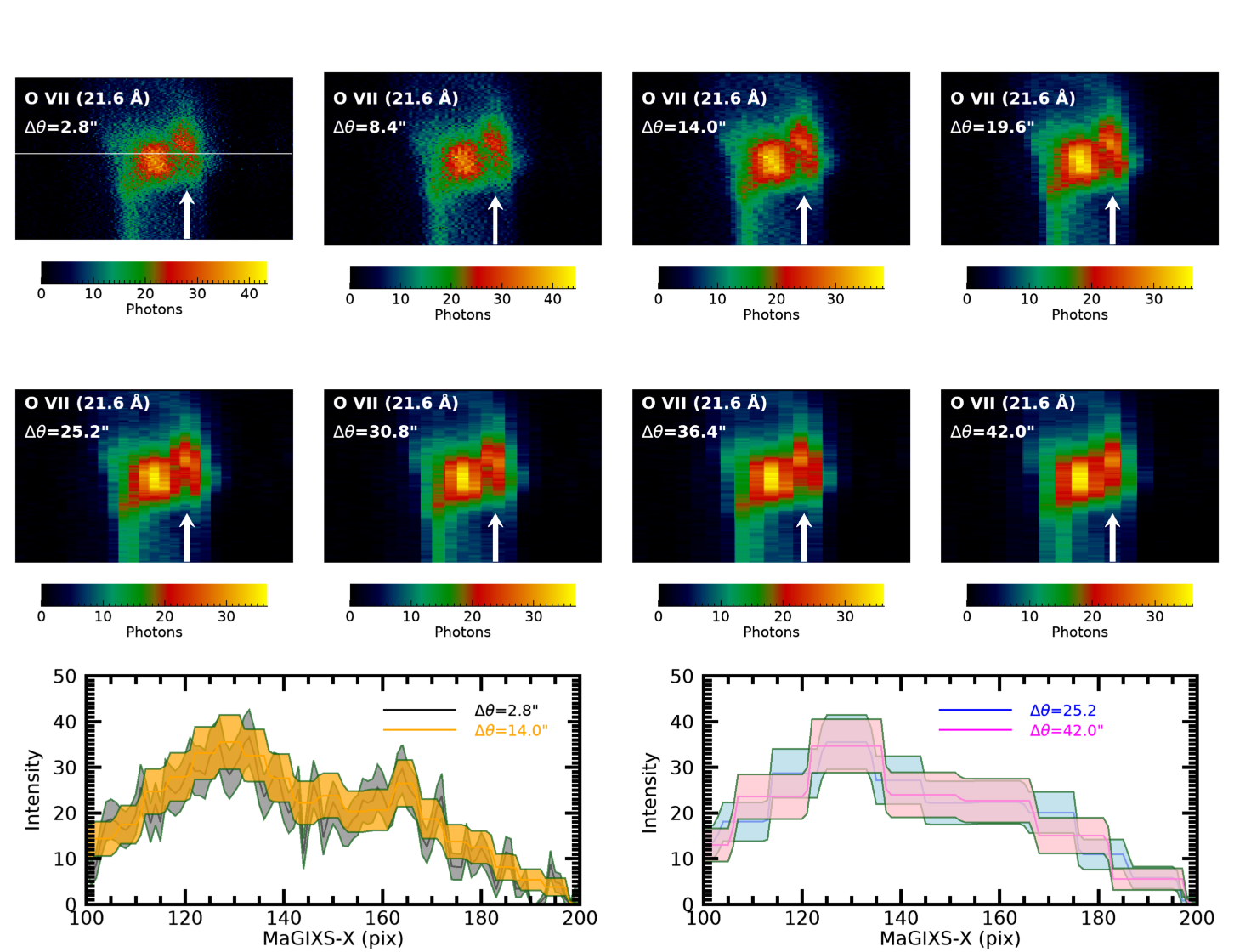}
    \caption{Comparison of O {\sc vii} maps inverted with different ${\delta}{\theta}$ values for a fixed ${\alpha}$ = 5${\times}10^{-5}$ and ${\rho}$ = 0.1.  Inversion with ${\delta}{\theta}$ $<$ 5 (Top) shows  two X-ray bright points observed distinctly, while ${\delta}{\theta}$ = 1, 2 over resolve the spatial structures, resulting in a ragged intensity distribution (bottom left), plotted with 1 ${\sigma}$ Poisson uncertainty (color filled section). (Middle) Inversion with ${\delta}{\theta}$ ${\geq}$ 5 indicate under resolving of structures where the two bright points started to merge, resulting in a smoothed intensity distribution (bottom right), plotted with 1 ${\sigma}$ Poisson uncertainty (color filled section).}
    \label{fig:dtheta_variation}
\end{figure}

The field angle grid size (${\delta}{\theta}$) on which to calculate the inversion, however, is more difficult to discern. The instrument is defined by a plate scale (size of a pixel in arcseconds) and a spatial resolution (size of the instrumental point spread function).  The spatial resolution must be at least twice the plate scale, but can be much larger than the plate scale.  In the case of \magixsone{}, there are multiple confounding factors.  First, the plate scale in the dispersion direction varies as a function of wavelength from 5.5\arcsec\ to
9\arcsec\; whereas the plate scale in cross-dispersion direction is 2.8\arcsec.  Secondly, the point spread function of \magixsone{} (${\sim}$ 27\arcsec) is significantly larger than these plate scales. For the study described  in this paper, we chose a grid size ${\delta}{\theta}$ = 8.4\arcsec\, but conceivably this value could have chosen a smaller or larger value.

Figure \ref{fig:dtheta_variation} shows the comparison of inverted spectrally pure images of O {\sc vii} at 21.6\,\AA, for different ${\delta}{\theta}$ values ranging from 2.8\arcsec\, to
42.0\arcsec). For ${\delta}{\theta}$ ${\leq}$ 14.0\arcsec\ (top panel) the two X-ray bright points can be observed distinctly with less cross-talk between the structures (indicated by the white arrow). For ${\delta}{\theta}$ $>$
14.0\arcsec, the bright point structures slowly start to merge (indicated by the white arrow), which denotes under resolution. For ${\delta}{\theta}$ $<$
8.4\arcsec, we observe high frequency structures in the spectrally pure images, as seen in the ragged intensity distribution (bottom left panel) across the structures marked by horizontal white line in top panel. The intensity variations are beyond 1 ${\sigma}$ Poisson statistics.  The intensity distribution across the structures show a very smoothed profile with reduced intensity, indicating under-resolving of the structures, as seen in the bottom right panel. We find the field angle grid size for \magixsone{} data that yield similar consistent results is 5.6\arcsec{} ${\leq}$ ${\delta}{\theta}$ ${\leq}$ 14\arcsec. This upper limit of optimum ${\delta}{\theta}$ is very close to half the value of \magixsone{} point spread function (27\arcsec). We report that by this exercise we do not find a single optimal ${\delta}{\theta}$ value for the inversion. Instead, we offer the guidance to choose a spatial resolution between the plate scale  of the instrument and the full width at half maximum (FWHM) of the PSF divided by 2.

 This paper serves as an outline for the methodology that must be completed for each new mission or study including spectroheliogram data. In this paper, we have determined the optimal values for ${\alpha}$ and ${\rho}$ for the \magixsone{} flight data.  However, we stress these choices are not universal and must be re-determined for each new instrument that is flown. Similarly, the choices for the ranges and resolutions of Log T and the field angles are instrument dependent and must be evaluated based on each instrument's sensitivity to temperature and spatial information. The work presented in this paper will be very helpful for next generation missions such as the Cubesat Imaging X-ray Solar Spectrometer (CubIXSS; \citealt{CubIXSS2021}) and potential small explorer (currently in Phase A) the EUV CME and Coronal Connectivity Observatory (ECCCO; \citealt{kathy2022}), which will carry overlappogram instruments. To facilitate inversion of data from future instruments and aid in instrument design, we have created an `overlappogram' package and made it publicly available in GitHub repository \footnote{https://eccco-mission.github.io/overlappogram/}.

 \section{Acknowledgements}
 \begin{acknowledgements}
We acknowledge the Marshall Grazing Incidence X-ray Spectrometer (\magixs) instrument team for making the data available through the 2014 NASA Heliophysics Technology and Instrument Development for Science (HTIDS) Low Cost Access to Space (LCAS) program, funded via grant NNM15AA15C. The authors thank and acknowledge Drs. Dan Seaton and Katharine Reeves, for their insightful comments.
\end{acknowledgements}

\bibliography{sample,sample631}{}
\bibliographystyle{aasjournal}



\end{document}